\documentclass[aps,pra,showpacs,superscriptaddress,floatfix,nofootinbib]{revtex4} 

\usepackage{amsfonts}
\usepackage{amsmath}
\usepackage{amsthm}
\usepackage{amscd}
\usepackage{amssymb}
\usepackage{amsxtra}
\usepackage{bm}           
\usepackage{bbm}
\usepackage{exscale}
\usepackage{epic,rotating}
\usepackage{float}
\usepackage{graphics,color,dsfont}
\usepackage{latexsym}
\usepackage{enumerate}
\usepackage{dcolumn}      
\usepackage{pstricks,pst-grad,fancybox,graphics}
\usepackage[scanall]{psfrag}
\usepackage[dvips]{epsfig}

\newcommand{\eins}{\ensuremath{\mathbbm 1}}

\newcommand{\bear}{\begin{eqnarray}}
\newcommand{\eear}{\end{eqnarray}}
\newcommand{\bearn}{\begin{eqnarray*}}
\newcommand{\eearn}{\end{eqnarray*}}
\newcommand{\kommentar}[1]{}

\newcommand{\mean}[1]{\ensuremath{\langle #1 \rangle}}

\newcommand{\bc}{\begin{center}}
\newcommand{\ec}{\end{center}}

\def\bi#1\ei {\begin{itemize}#1\end{itemize}}
\def\bea#1\eea {\begin{align}#1\end{align}}
\def\bean#1\eean {\begin{align*}#1\end{align*}}
\def\ben#1\een {\begin{equation*}#1\end{equation*}}
\def\be#1\ee {\begin{equation}#1\end{equation}}
\def\bes#1\ees {\begin{equation}\begin{split}#1\end{split}\end{equation}}

\newcommand{\halbe}{\frac{1}{2}}

\newcommand{\sz}[1]{\sigma^z_{#1}}
\newcommand{\sx}[1]{\sigma^x_{#1}}
\newcommand{\sy}[1]{\sigma^y_{#1}}

\begin{document}

\title{Local renormalization method for random systems}

\author{O.~Gittsovich}
\affiliation{Institut f\"ur Theoretische Physik, Universit\"at Innsbruck, Technikerstra{\ss}e 25,6020 Innsbruck, Austria}
\affiliation{Institut f\"ur Quantenoptik und Quanteninformation, ~\"Osterreichische Akademie der Wissenschaften, Technikerstra{\ss}e 21a, 6020 Innsbruck, Austria}

\author{R.~H{\"u}bener }
\affiliation{Institut f\"ur Theoretische Physik, Universit\"at Innsbruck, Technikerstra{\ss}e 25,6020 Innsbruck, Austria}

\author{E.~Rico}
\affiliation{Fakult\"at  f\"ur Physik, Universit\"at Wien, Boltzmanngasse 5, A-1090 Vienna, Austria.}

\author{H.~J.~Briegel}
\affiliation{Institut f\"ur Theoretische Physik, Universit\"at Innsbruck, Technikerstra{\ss}e 25,6020 Innsbruck, Austria}
\affiliation{Institut f\"ur Quantenoptik und Quanteninformation, ~\"Osterreichische Akademie der Wissenschaften, Technikerstra{\ss}e 21a, 6020 Innsbruck, Austria}

\begin{abstract}
In this paper we introduce a real-space renormalization transformation for random spin systems on 2D lattices. The general method is formulated for random systems and results from merging two well known real space renormalization techniques, namely the strong disorder renormalization technique (SDRT) and the contractor renormalization (CORE). We analyze the performance of the method on the 2D random transverse field Ising model (RTFIM).
\end{abstract}

\date{\today}

\maketitle

\section{Introduction}

Most physical systems are disordered and the description and modeling of such systems is one of the most special problems in condensed matter physics. In the early 70's the role of the disorder in physical systems was discussed in several papers. Harris \cite{Harris:1974dn} formulated a criterion for the relevance of the weak disorder caused by locally random impurities in the system. According to the criterion, the relevance of the disorder depends on the sign of the critical exponent for the specific heat. Just one year later, Imry and Ma \cite{Imry:1975rw} formulated another criterion, which points out the relevance of the weak disorder in less than four dimensions (the ordered state became unstable against an arbitrarily weak random field). As it became clear later these criteria can be understood in terms of the real space renormalization group (see for instance \cite{Cardy:2000fu,Refael:2004pi} and references therein).

Real space renormalization group (RG) methods for quantum disordered systems were applied first in the late of 70's. In the pioneering work on this topic Ma, Dasgupta and Hu considered a spin-$1/2$ antiferromagnetic Heisenberg chain, where the coupling strengths were assumed to be stochastically distributed \cite{Ma:1979hl}. The authors studied the model at zero temperature using a method that essentially relies on the reduction of the number of degrees of freedom in the system. The method attracted a lot of attention, was extensively studied and developed further by Fisher \cite{Fisher:1992qd,Fisher:1995qr} and has been used to investigate a big variety of systems in one and two dimensions (a review on the real space RG approach can be found in \cite{Igloi:2005fr}). The method has recently been referred to as strong disorder renormalization technique (SDRT). 

Indeed it turned out that the behavior of a system with randomness is in many cases quite different from the non-random case. The main ingredient of the disordered systems, which has no counterpart in the systems without disorder, is the existence of so called rare regions, i.e., regions possessing atypical properties (for the phase under consideration) compared to the rest of the system. It is known that this kind of rare effects can govern the behavior of the systems at long distances and result in exotic phases, \emph{e.g.} Griffiths-McCoy \cite{Griffiths:1969kx,McCoy:1969yq,Pich:1998mz,Motrunich:2000hl,Igloi:2005fr} phase. It is worth to note that the application of the method to two dimensional systems is not straightforward, since it distorts the geometry of the underlying lattice, and only numerical calculations are possible. Therefore analytic proofs like asymptotic exactness of the SDRT in thermodynamical limit (see \cite{Fisher:1992qd,Fisher:1995qr}) do not apply. We will come back to the SDRT in the next section and discuss it in more detail.

Despite its beauty, the SDRT is a perturbative method and applying it to the finite sized systems may cause a problem. A non-perturbative RG method, that for our knowledge has not been applied to random systems yet, was introduced by Morningstar and Weinstein in \cite{Morningstar:1996oq}. This method is called contractor renormalization (CORE) group approach and is especially suited for lattice systems (for CORE applications see \cite{Altman:2002la,Berg:2003yq,Budnik:2004ys,Capponi:2004rt,Siu:2008vn}). By definition, this method keeps the eigenvalues of the low energy sector and produces an optimal truncation operator from the original Hilbert space to the effective one. In other words, the CORE is a non-perturbative block-spin renormalization, which uses exact diagonalization to extract the effective interactions in a coarse grained system.

Having a non-perturbative method on the one hand and ideas of spatially local renormalization of the system from the SDRT on the other hand, we introduce a method that unifies both techniques and is suited to investigate two dimensional disordered systems. The purpose of this paper is to show that such merging is possible and results in a non-perturbative real space renormalization transformation for 2D quantum systems at zero temperature that preserves the underlying lattice geometry.

A reliable real space renormalization group method for description of the low temperature behavior of some system gives information about the long distance properties of the system while keeping the fundamental structure of the ground state. This fact is especially relevant in quantum random systems where the entanglement properties of the ground state have been identified as the key feature in understanding the behavior of these materials \cite{Ancona-Torres:2008lq, Fidkowski:2008fp}.

Since the method we are about to introduce involves local real-space renormalization steps we will have to analyze errors introduced by these local operations. As a benchmark we are going to use statistical arguments, showing that long range interactions are not important in the renormalized system and therefore can be neglected, i.e., if we consider a model that initially has only nearest neighbors interactions, we can neglect next nearest or more sophisticated terms introduced by renormalization and proceed further with a model that has only nearest neighbor interactions. The statistical justification of the fact that our method can be applied locally in the real space and without renormalizing the whole lattice at once, is a crucial point of this paper.

The paper is organized as follows. In the next section we are going to discuss two important real space renormalization techniques in more detail and provide the idea of constructing a novel method for renormalization of disordered spins systems on 2D rectangular lattices. In section III. we analyze the performance of the introduced renormalization transformation and consider several toy models to prove the negligibility of the long range spin-spin interactions that might appear during the renormalization process. We give some open problems and incitations for future investigations in outlook section.

\section{Real space renormalization group methods and random systems.}

\subsection{Strong Disorder Renormalization Technique}
The name strong disorder renormalization technique reveals perfectly the idea of real space RG for random systems introduced by Ma, Dasgupta and Hu in \cite{Ma:1979hl}.

There are {\it a priori} several different situations that can appear in disordered systems in the thermodynamical limit. When the size of the system increases and the effective disorder becomes a major effect compared to the thermal or to the quantum fluctuations, this effective disorder can either become

\begin{itemize}
\item smaller and smaller without bound: the system is then controlled by a pure fixed point,
\item larger and larger without bound: the system is then controlled by an infinite disorder or infinite-randomness fixed point (IRFP),
\item or it may converge towards a finite level: the system is then controlled by a finite disorder fixed point.
\end{itemize}

A class of systems whose critical behavior is governed by an infinite-randomness fixed point (IRFP) is characterized by a very broad distribution of couplings and a dynamical exponent $z$ that becomes infinite at the critical point. In certain models, any initial disorder, even very small, drives the system towards the IRFP at the large scale: in particular, this is the case for the random antiferromagnetic quantum spin chain with local spin systems with $S=\frac{1}{2}$ (see also \cite{Fisher:1992qd}).

We will illustrate very briefly a concrete scheme for renormalization of systems with infinitely strong disorder in one dimension on example of the random transverse field quantum Ising chain (RTFIC) (for detailed consideration see \cite{Fisher:1995qr}). The system has the following Hamiltonian
\be
H=-\sum_{(ij)} J_{ij} \sigma^z_i \sigma^z_j - \sum_i h_i \sigma^x_i.
\label{HIsing}
\ee
The basic strategy is to find the strongest coupling in the chain (it can be either $\{J_{ij}\}$ or $\{h_i\}$) and minimize the corresponding term in the Hamiltonian. The degrees of freedom associated with this maximum energy scale $\Omega_0=\max\{J_{ij},h_i\}$ are then frozen at lower energy scales.

If the strongest coupling is a field, say $h_k$ then the spin $\sigma_k$ is put in its local ground state, i.e., in the $x$-direction, causing it to become non-magnetic. Effective interactions are then generated between its nearest neighbors; but, as all other nearby couplings are likely to be much smaller than $h_i$, these can be treated by second order perturbation theory. This introduces new effective interactions
\be
\tilde{J}_{ij} \simeq \frac{J_{ik}J_{kj}}{h_k},
\ee
where $i$ and $j$ are the nearest neighbors of the $k$.

If the strongest coupling is an interaction, say $J_{kl}$, then two spins are combined forming a cluster which, in the zeroth order of perturbation theory, has a double degenerate ground state (both up or both down) and thus can be represented again by an effective two-level particle: a new spin. The effective local magnetic field being applied to the cluster $(kl)$ is
\be
\tilde{h}_{(kl)} \simeq \frac{h_k h_l}{J_{kl}},
\ee
which results from the second order perturbation theory, where magnetic fields, acting on two spins are considered to be small.

The magnetization of the cluster will be the sum of magnetizations of single spins $k$ and $l$, i.e., it changes additively $m_{(kl)}=m_k + m_l$. Since all new couplings are smaller than the initial one $\Omega_0$, the energy is rescaled and the maximum energy is reduced (for more details see \cite{Fisher:1992qd,Fisher:1995qr}). Note that the decimation as described above would change the geometry of the system in dimensions higher than $D=1$, so that we consider the spins to be the vertices of a somewhat random graph with the RG modifying the spatial structure in these larger dimensions \cite{Igloi:2005fr, Motrunich:2000hl}.

If the quantum disordered phase is renormalized, the fields eventually tend to dominate the bonds and at small values of $\Omega$ almost all decimations are cluster annihilations and the effective interactions connecting them becoming weaker and weaker; in the procedure of the renormalization, the system hence becomes a collection of asymptotically uncoupled clusters with a broad distribution of effective fields. In the ordered phase, in contrast, the interactions tend to dominate the fields at low energies, and most decimations are thus decimations of bonds; eventually this causes an infinite cluster to form. The zero temperature quantum transition between these phases is a novel kind of percolation with the annihilation and aggregation of clusters competing at all energies at the critical point \cite{Igloi:2005fr, Motrunich:2000hl,Ma:1979hl, Fisher:1992qd, Fisher:1995qr}.

Before closing this section we would like to point out that the SDRT consists of successive local renormalizations in the real space, where no long range interactions are considered. It means that after an elementary renormalization step the system is described by a Hamiltonian with only nearest neighbors interactions (if one had started with a nearest neighbors interactions Hamiltonian) and no next nearest neighbors appear in the Hamiltonian. Strictly speaking, after every RG step the ground state of the effective Hamiltonian will deviate from the ground state of the initial one, but the error will become asymptotically small in the thermodynamical limit as has been proven by Fisher in \cite{Fisher:1995qr}. This is the feature we want to retain in our ansatz later on.

\subsection{The CORE method}\label{coresection}

The CORE is the Hamiltonian version of the Kadanoff-Wilson real space RG transformation for lattice field theories and lattice spin systems and relies on contraction and cluster expansion techniques. We briefly sketch the main idea of the CORE and how it works and refer to ref. \cite{Morningstar:1996oq} for details.

The first step in this method is to choose small clusters, elementary blocks which cover the lattice. After that, one picks up some of the clusters (since the CORE was introduced for systems with no disorder and with translation symmetry, all clusters are the same) and considers the part of the whole Hamiltonian that corresponds to this cluster. In what follows we call this part of the Hamiltonian cluster Hamiltonian. For the cluster Hamiltonian one has to choose states that are relevant for the description of physical behavior of the cluster (the number and the form of these states can vary depending on the particular model). The span of the chosen states forms the effective Hilbert space of the cluster. Then a projection $P_{eff}$ on the effective Hilbert space of the cluster is constructed. This projection is used to obtain the so-called range-1 terms of the Hamiltonian expansion ($h_i^{(1)}=P_{eff}H^i_{cluster}P_{eff}$). The range-2 terms arise from the Hamiltonian that corresponds to two adjacent (connected) clusters. The states of the effective Hilbert space of the connected clusters is obtained by taking tensor products of the states single clusters. Afterwards a unitary matrix is constructed by means of which the range-2 terms are produced (this matrix is called triangulation matrix \cite{Morningstar:1996oq}). This procedure is iterated to achieve the range-$N$ terms. Finally the expansion of the truncated Hamiltonian, which is the effective Hamiltonian after single renormalization step is written as
\be
H_{\text{eff}}= \sum_i h_i^{(1)}+\sum_{<i,j>} h_{i,j}^{(2)} + \sum_{<i,j,k>}  h_{i,j,k}^{(3)}+ \cdots
\label{expansion}
\ee
where $h_{i_1,\dots i_N}^{(N)}$ stands for range-$N$ term. For more details and rigorous proof that the truncated Hamiltonian can be expanded in this way, we refer the reader to ref. \cite{Morningstar:1996oq}.

Note that for the construction of the range-$N$ terms in the expansion (\ref{expansion}) one obtains the eigenvalues $\{\epsilon_n\}$ and eigenvectors $\{|n\rangle\}$ by the exact diagonalization of $N$ contiguous clusters. The optimal truncation operator (triangulation matrix) is obtained by a Gram-Schmidt orthogonalization of the eigenvectors of the Hamiltonian projected on the effective Hilbert space. In this way, a basis $\{|\phi_n\rangle \}$ (the remnant eigenstates of the range-$N$ Hamiltonian) is built such that the first vector overlaps with the lowest energy eigenvector and those above, the second one with the second lowest and those above and so on, i.e.,
\be
|\phi^N_n\rangle=\sum_{m \ge n} \lambda_m |m\rangle.
\ee
In fact, this reduced basis stems from the $QR$-decomposition of the overlap matrix between the reduced Hilbert space and the space of exact eigenvectors of the complete Hamiltonian\cite{Siu:2008vn}.

Usually, two situations can occur after several steps of the renormalization. The Hamiltonian either flows to a point where it can be solved exactly and the correlation length in the effective lattice model goes to zero, or the system is self-similar at every scale, the correlation length diverges and the mass gap goes to zero: at this point, the system is said to be at the critical point.

In summary, the CORE has two major advantages over traditional perturbative real-space renormalization schemes:

\begin{itemize}
\item the CORE is not an expansion in weak/strong bonds between block-spins. Its convergence does not necessarily depend on the existence of a large gap to the discarded states of the Hilbert space.
\item the CORE is based on an exact mapping form the original Hamiltonian to an effective Hamiltonian, whose truncation error can be estimated numerically by calculating higher orders in the expansion.
\end{itemize}

Finishing this section we point out that when the Hilbert space dimension is reduced, the CORE provides a good description of the initial states in terms of the renormalized states. In order to estimate the quality of the description of the states from the constructed effective Hilbert space, one can use an overlap of the lowest energy states $|m \rangle$ with the remnant states $|\phi^N_m\rangle$, when the range-$N$ term in the expansion is constructed. Note that both issues are related, as the closer the truncated space is to the exact one, the smaller is the number of terms that should be kept in the cluster expansion for a given error.

\subsection{Combining the CORE and the SDRT.}

In this section we provide the idea how to construct a real space renormalization method for two dimensional disordered systems. The details concerning the accuracy of the method are presented in section \ref{lrnonun}.

\subsubsection{General idea of the method.}\label{gen_idea}
The real space RG method we are about to introduce combines the SDRT to target the clusters to be decimated as the ones with the biggest energy gaps and the CORE as a tool to obtain the effective dynamics at a new scale. 

\begin{figure}[!ht]
\includegraphics[width=0.4\columnwidth]{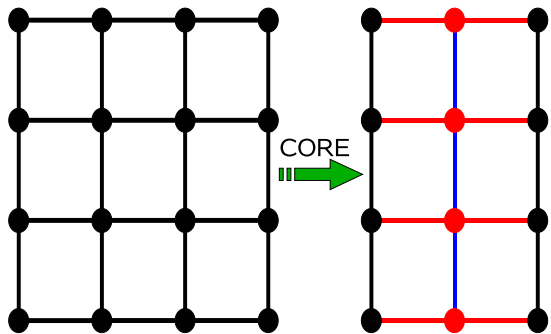}
\caption{(Color online) Renormalization of one ladder from 4x4 lattice. Red dots are new effective 2-level systems, which interact via effective couplings between each other (blue
dashed lines). Note that the ladder is not necessarily a column, it can be also a row of the initial lattice.}
\label{CORE4x4}
\end{figure}

In the Fig.\ref{CORE4x4} we briefly demonstrate how the renormalization works. The renormalization is applied to a ladder of spins which is a) traversing the whole lattice and b) contains the local two spin Hamiltonian with the biggest energy gap. Accordingly, the $4\times 4$ square lattice is transformed to become a $4\times 3$ square lattice with new effective particles and nearest neighbor interactions. The choice to renormalize a whole ladder is made to maintain the lattice geometry at every step of the transformation. Indeed the renormalization transformation, which involves a single ladder of the initial rectangular lattice, results again in a rectangular lattice and therefore preserves the initial geometry of the lattice. Note that the ladder involved in the renormalization transformation is either a column or a row of the lattice. In what follows we call the renormalization of the single ladder a \emph{renormalization step}.

The choice of the ladder follows the position of the local two spin Hamiltonian with the biggest gap. Once this Hamiltonian is found, the whole ladder is renormalized. The criterion of targeting the ladder is arbitrary, but might have an impact on the outcome of the procedure for some Hamiltonians. For example one can target the ladder, which contains a maximal number of local Hamiltonian with a rather big energy gap, albeit the local Hamiltonians with a maximum energy gap does not belong to the ladder. We leave the discussion of the different strategies of the ladder targeting as an open question.

\begin{figure}[!ht]
\includegraphics[width=0.5\columnwidth]{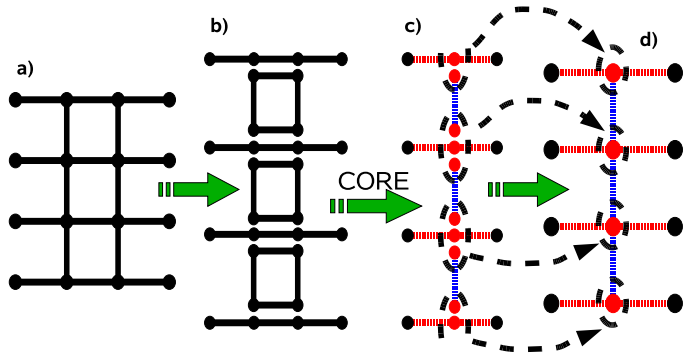}
\caption{(Color online) Four essential constituents of the single renormalization step: a) choice of the relevant ladder b) decomposition of the ladder into 4 particle terms: plaquettes and chains c) renormalization of each 4 particle term via CORE d) averaging of the local effective terms (dashed circles) and assembling of the effective hamiltonians into a renormalized lattice.}
\label{4steps}
\end{figure}

Every renormalization step is a sequence of two basic renormalization transformations. In order to see how these basic transformations enter the renormalization, we discuss the renormalization of the ladder in more details. First, the ladder is decomposed into four-spin blocks, such that some of the blocks form chains and some of them form plaquettes (from FIG.\ref{4steps} {\bf a)} to FIG.\ref{4steps} {\bf b)}). The chain terms correspond to the interaction of every rung (two spins) of the ladder to its nearest neighbors. The plaquette terms describe interactions between two rungs inside the ladder. Note that every pair of spins in the ladder (the rung) contributes to two plaquette terms and to one chain term.
After the decomposition, each term, representing one of the two basic lattice substructures is renormalized separately using the CORE. This leads to a set of effective 2 and 3 particle Hamiltonians (FIG. \ref{4steps} {\bf c)}). In the final step the effective Hamiltonians are assembled to the renormalized Hamiltonian on the smaller lattice (FIG. \ref{4steps} {\bf d)}).

The renormalization step can hence be summarized as follows:
\begin{enumerate}
\item Target the ladder with the biggest local energy gap.
\item Define the reduced Hilbert space by the lowest energy sector of every pair of spins in the ladder and the rest of the untouched spins in the lattice.
\item Compute exactly the eigenvalues of the four spin problem (the hardest computational step).
\item Obtain the Hamiltonian on the next scale and rescale the unit of distance and energy.
\item Iterate the procedure.
\end{enumerate}

It is noteworthy that step 2.\ and 3.\ of the algorithm rely on an unusual implementation of the CORE method. In the introductory part we mentioned that CORE makes use of a uniform blocking of the lattice (elementary blocks have the same form because of the translational symmetry and used to construct the range-1 terms of the Hamiltonian expansion). Since we now perform the renormalization transformation locally (translational symmetry does not apply in the presence of randomness), we need to introduce a non-uniform blocking. This non-uniform blocking will be demonstrated using the example of the renormalization of four spins in a chain configuration. The performance of such blocking will be analyzed in the next section. 
\subsubsection{Elementary steps for the successive renormalization transformation.}\label{elemsteps}
The elementary renormalization transformations of the four spin terms mentioned above can be divided into two groups.

\begin{figure}[!ht]
\includegraphics[width=0.25\columnwidth]{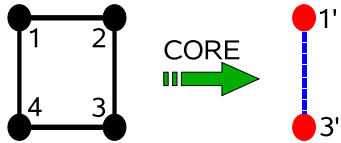}
\caption{(Color online) Renormalization of four spins in a plaquette configuration. The renormalization transformation results in two new particles $1^{\prime}$ and $3^{\prime}$ and an interaction between them.}
\label{COREplaq}
\end{figure}

The first type is renormalization of a plaquette, that results in two new particles and new interaction between them (Fig. \ref{COREplaq}). We use the CORE to renormalize spins in the plaquette configuration. Each pair forms an elementary cluster and used to construct the range-1 term. The interaction between two effective particles is given by the range-2 term.

\begin{figure}[!ht]
\includegraphics[width=0.4\columnwidth]{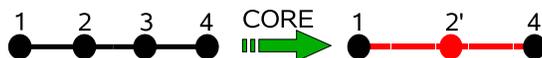}
\caption{(Color online) Renormalization of four spins in a chain configuration. The renormalization transformation results in three particles $1$, $2^{\prime}$ and $4$. The interaction between the particles is of the short range character (only nearest neighbors interact).}
\label{COREchain}
\end{figure}

The second type is a renormalization of 4 spins in a chain configuration (FIG. \ref{COREchain}). The latter introduces effective interactions to the neighboring spins that increase the accuracy of the method. To use the CORE as it described in FIG. \ref{COREchain} we need to modify it. That is to say the size of the elementary clusters varies. We call this way of implementation of the CORE - non-uniform blocking. Two central spins form an elementary cluster as well as boundary spins. Constructing the range-1 terms the initial Hilbert space of the boundary spins is kept, whereas the effective Hilbert space of the central two spins is spanned by the ground state and by the first excited state of the two spin Hamiltonian. The range-2 terms are achieved by constructing the triangulation matrix, while the effective Hilbert space is a tensor product of the Hilbert spaces of two boundary spins and the span of the two lowest eigenstates of the Hamiltonian of two central spins. This modification reflects the fact that the renormalization transformation is local in real space due to intrinsic disorder of the system. We will come back to the non-uniform blocking and analyze its performance numerically in the next section.
%


\section{Estimation of long distance and multi-spin interactions}\label{lrnonun}

In order to investigate the performance of the elementary renormalization transformations described briefly in section \ref{elemsteps} we pick up a particular model that is a 2D random transverse field quantum Ising model (RTFIM). The Hamiltonian of the 2D RTFIM possesses $\mathbb{Z}_2$-symmetry that can be exploited in the renormalization transformation and provides a special form of the effective Hamiltonian after each renormalization step (the same observations were made for the 1D Ising model in \cite{Morningstar:1996oq}). 

\subsection{$\mathbb{Z}_2$-symmetry of the 2D Random Transverse Field Ising Model}\label{symmarg}

The 2D RTFIM is described by the Hamiltonian
\be
H=-\sum_{(ij)} J_{ij} \sigma^z_i \sigma^z_j - \sum_i h_i \sigma^x_i
\ee
where $\{J_{ij} \}$ are random interactions and the random transverse fields $\{h_i\}$ leading to the quantum fluctuations. The specific form of the distribution will be defined later.

This Hamiltonian is invariant under the transformation $\sigma^z_i \to  - \sigma^z_i$ ($\mathbb{Z}_2$-symmetry). The CORE has to preserve this symmetry so that the most general form the renormalized Hamiltonian can take is
\be
H_{\text{eff}}= - \sum_{\{\mu\}, i} g_{\{\mu\}} \hat{O}^{\{\mu\}}_i, \hspace{3ex} \hat{O}^{\{\mu\}}_i = \sigma^{\mu_1}_i \sigma^{\mu_2}_{i+1} \cdots \sigma^{\mu_n}_{i+n}
\ee
where $i$ is the site index, $\{\mu\}=\{\mu_1,\dots,\mu_n\}$ is the multi-index ($\mu_i\in\{u,x,y,z\}$) and the $g_{\{\mu\}}$'s are the couplings. 

Due to the $\mathbb{Z}_2$-symmetry of the model the only operators that can appear in the one particle Hamiltonian in the cluster expansion are $\{\sigma^0, \sigma^x \}$; in the two particle nearest neighbor interactions, the symmetries allow terms of the form $\{ \sigma^z \sigma^z \}$ from the Ising interaction and also $\{\sigma^x \sigma^x, \sigma^y \sigma^y \}$ and the only three site operators that can appear are: $\{\sigma^x \sigma^x \sigma^x, \sigma^x \sigma^z \sigma^z, \sigma^z \sigma^z \sigma^x, \sigma^z \sigma^x \sigma^z, \sigma^x \sigma^y \sigma^y, \sigma^y \sigma^y \sigma^x,$ $\sigma^y \sigma^x \sigma^y , \sigma^x \sigma^0 \sigma^x, \sigma^y \sigma^0 \sigma^y, \sigma^z \sigma^0 \sigma^z \}$. From the above discussion we conclude that the $\mathbb{Z}_2$-symmetry puts certain constrains on the form of the range-$N$ terms that can appear in the expansion of the effective Hamiltonian (\ref{expansion}).

Exploiting the symmetry arguments we will investigate the relevance of the range-3 and range-4 terms that remain in the expansion (\ref{expansion}), when the renormalization follows the $\mathbb{Z}_2$-symmetry
\be
H^{Ising}_{\text{eff}}= \sum_i h_i^{(1)}+\sum_{<i,j>} h_{i,j}^{(2)} + \sum_{<i,j,k>}  h_{i,j,k}^{(3)}+ \cdots
\nonumber
\ee

To achieve the task we will consider several scenarios of non-uniform and uniform blocking in various toy models.

\subsection{Chain of four spins}

\begin{figure}[!ht]
\includegraphics[width=0.35\columnwidth]{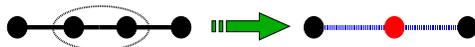}
\caption{(Color online) Renormalization of four spins of the RTFIM in a chain configuration using a non uniform blocking. The encircled pair of spins and spins on the boundaries of the chain are used to form range-1 terms for the effective Hamiltonian. The circles on the right hand side of the figure correspond to the range-1 terms in the effective Hamiltonian. These circles are connected by the lines that correspond to range-2 terms.}
\label{4to3}
\end{figure}

First of all we consider a chain of four spins, which after renormalization becomes a chain of three spins (FIG. \ref{4to3}). (This step is an essential part of renormalization transformation as discussed in section \ref{elemsteps}). The encircled pair of spins and spins on the boundaries of the chain form the range-1 Hamiltonians. The effective Hamiltonian consists of range-1, -2, and -3 terms.

Our goal here is to estimate the range-3 terms, which appear in the effective Hamiltonian.
\begin{figure}[!ht]
\includegraphics[width=0.45\columnwidth]{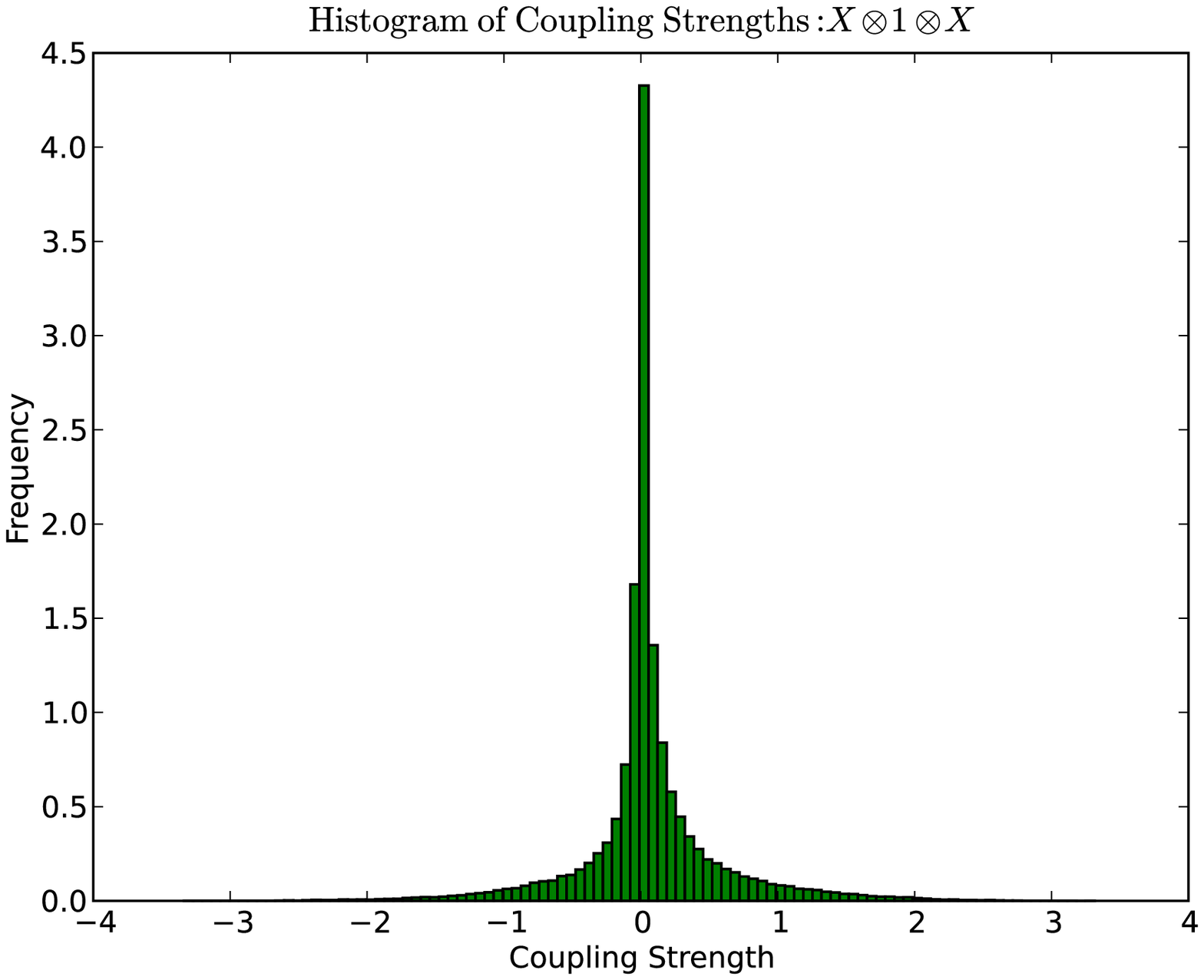}
\includegraphics[width=0.45\columnwidth]{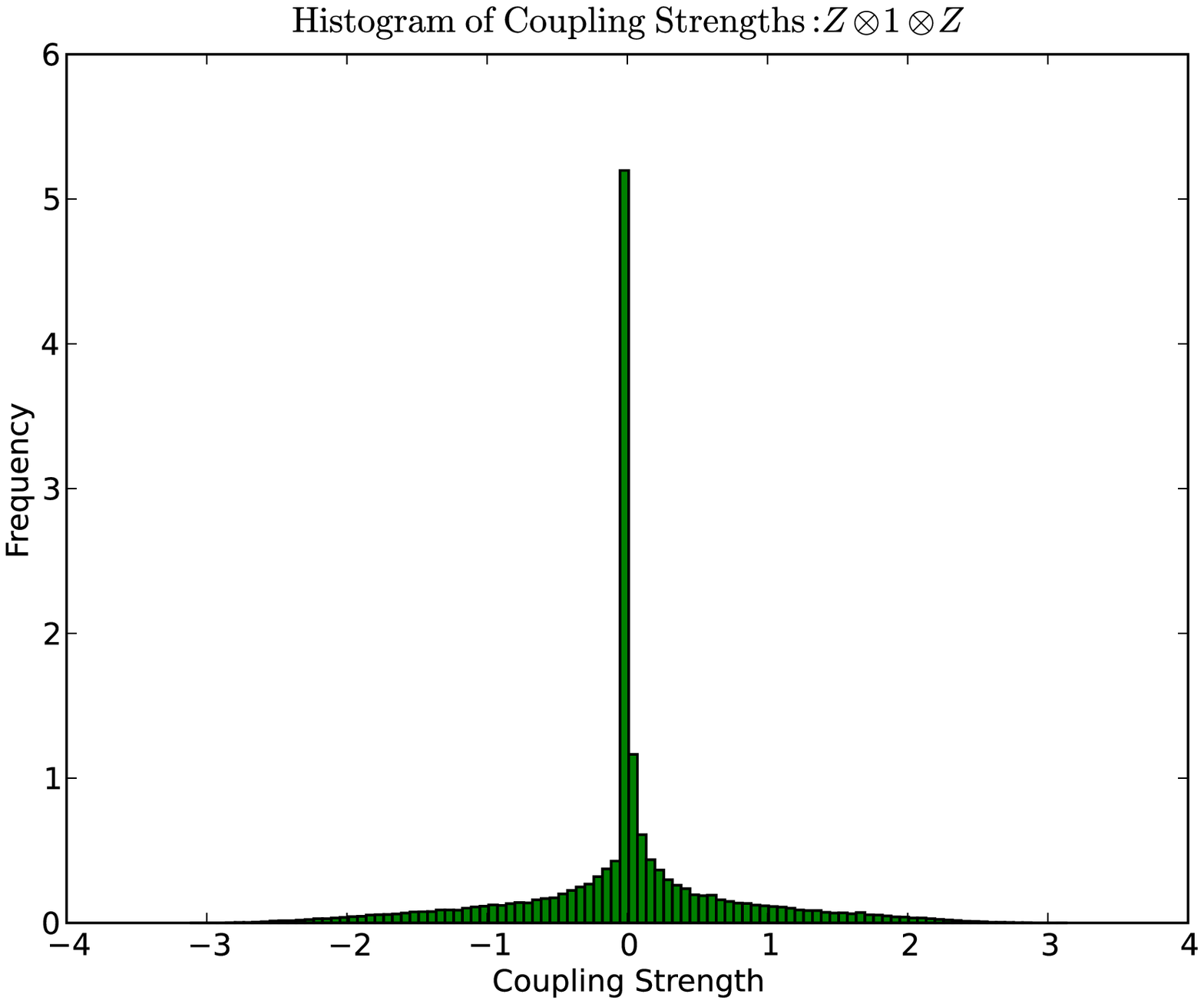}
\caption{(Color online) Renormalization of four spins of the RTFIM in a chain configuration using a non-uniform blocking. (Left plot) $XX$ coupling between the first and the third particle. $\mean{\sigma_x\otimes\eins\otimes\sigma_x}=0.041$, $\sigma(\sigma_x\otimes\eins\otimes\sigma_x)=0.704$. (Right plot) $ZZ$ coupling between the first and the third particle. $\mean{\sigma_z\otimes\eins\otimes\sigma_z}=0.003$, $\sigma(\sigma_z\otimes\eins\otimes \sigma_z)=0.746$.}
\label{NNN4to3}
\end{figure}
There are 10 possible terms in the range-3 Hamiltonian that satisfy the $\mathbb{Z}_2$-symmetry (see section \ref{symmarg}). As our simulations show all this terms are negligibly small in the presence of disorder. In FIG. \ref{NNN4to3} we present the $XX$ (the upper picture) and $ZZ$ (the lower picture) couplings between the first and the third particle of the renormalized chain. The initial couplings were uniformly distributed on the interval $[0,1]$ and presented statistics were taken after testing $10^5$ different configurations. As one can see from the FIG. \ref{NNN4to3} the resulting distributions of both $XX$ and $ZZ$ couplings are symmetric and centered at 0. The standard deviations are $0.704$ and $0.746$ for $XX$ and $ZZ$ interactions respectively.

\subsection{Ladder of six spins. Uniform blocking}
\begin{figure}[!ht]
\includegraphics[width=0.3\columnwidth]{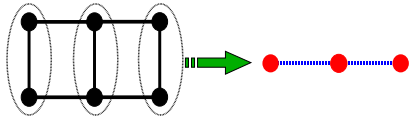}
\caption{(Color online) Renormalization of six spins of the RTFIM using a uniform blocking. To form range-1 terms of the effective Hamiltonian the encircled pairs of spins are used. The circles on the right hand side of the figure correspond to the range-1 terms in the effective Hamiltonian. These circles are connected by the lines, that correspond to range-2 terms.}
\label{6to3}
\end{figure}

In our next example we consider a ladder of six spins, that we transform to a chain of three spins using the uniform blocking (FIG. \ref{6to3}). The encircled pairs of spins are taken to form range-1 terms in the expansion of the effective Hamiltonian. As in the previous example the expansion will comprise up to range-3 terms.

\begin{figure}[!ht]
\includegraphics[width=0.45\columnwidth]{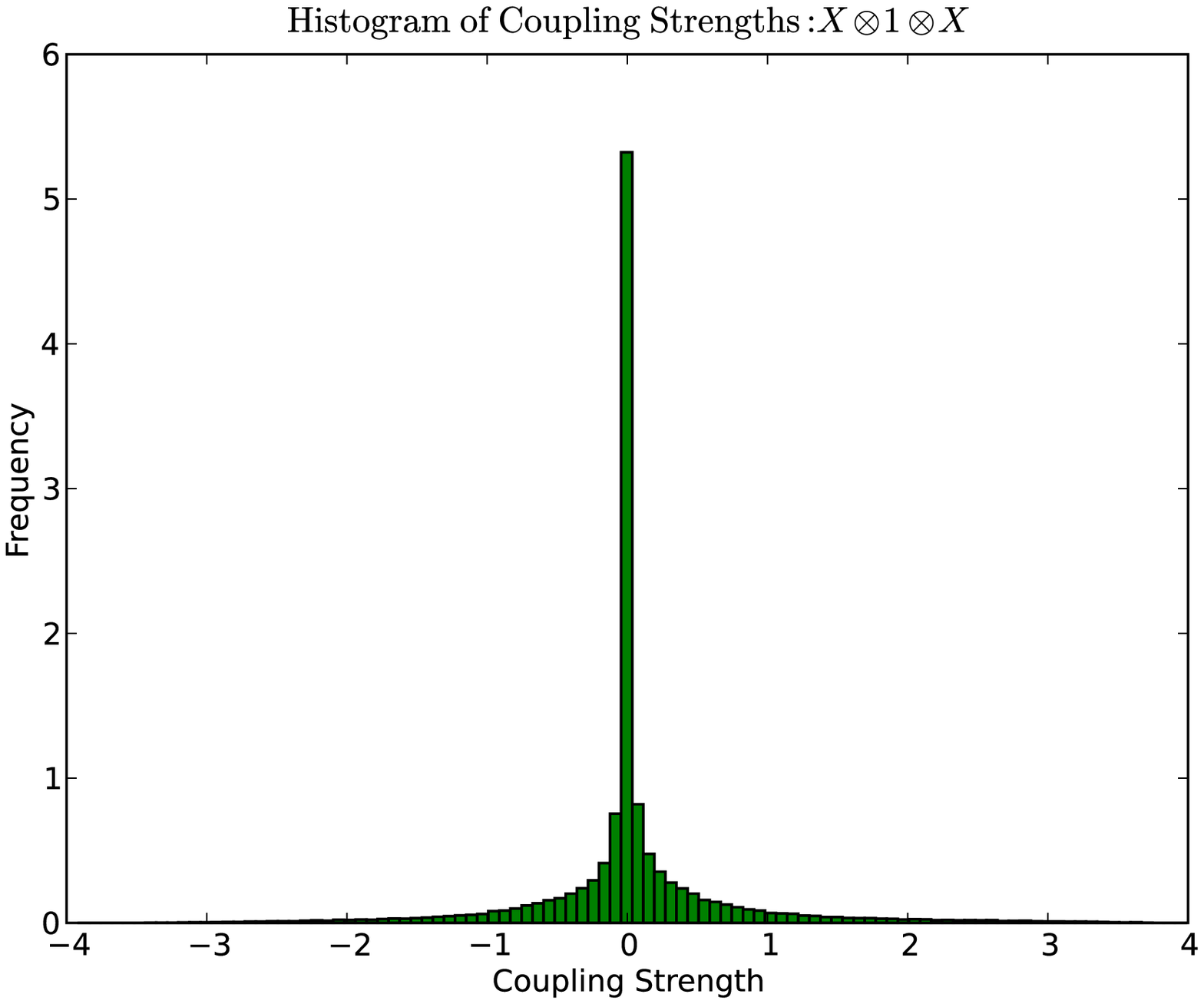}
\includegraphics[width=0.45\columnwidth]{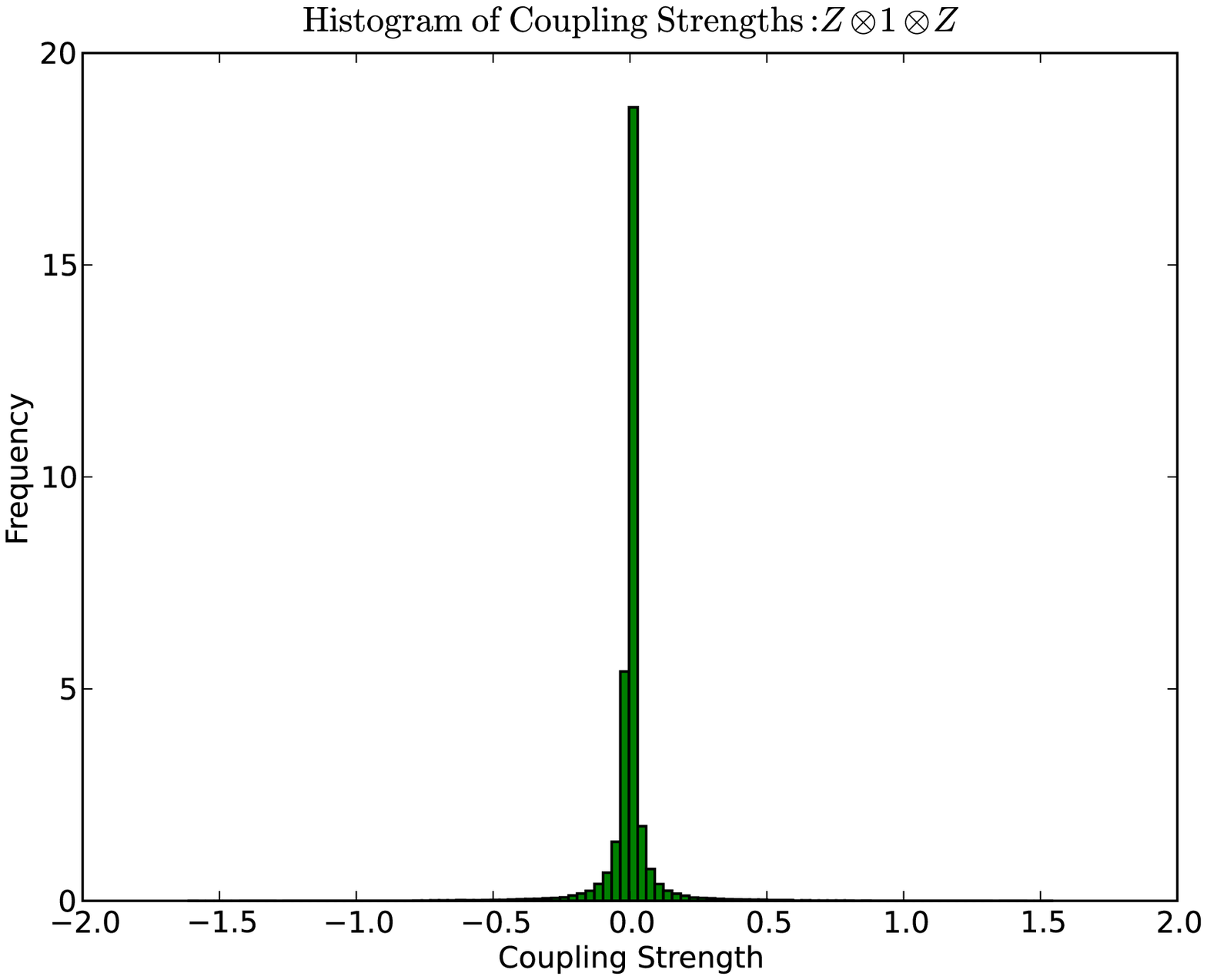}
\caption{(Color online) Renormalization of six spins of the RTFIM using a uniform blocking. (Upper plot) $XX$ coupling between the first and the third particle. $\mean{\sigma_x\otimes\eins\otimes \sigma_x}=0.064$, $\sigma(\sigma_x\otimes\eins\otimes\sigma_x)=0.500$. (Lower plot) $ZZ$ coupling between the first and the third particle. $\mean{\sigma_z\otimes\eins\otimes\sigma_z}=0.0003$, $\sigma(\sigma_z\otimes\eins\otimes\sigma_z)=0.0996$.}
\label{NNN6to3}
\end{figure}

In FIG. \ref{NNN6to3} we present the distributions of $XX$ and $ZZ$ coupling strengths between the first and the third particle in the resulting chain. The initial distribution was again a uniform distribution from the interval $[0,1]$ and we collected statistics after testing $10^5$ configurations.
As in the previous example both of the resulting distributions have a peak by 0 and standard deviations $0.500$ and $0.0996$ for $XX$ and $ZZ$ interactions respectively.

\subsection{Ladder of six spins. Non-uniform blocking}

\begin{figure}[!ht]
\includegraphics[width=0.3\columnwidth]{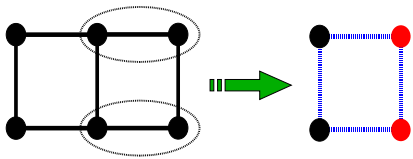}
\caption{(Color online) Renormalization of six spins of the RTFIM using a non-uniform blocking. To form range-1 terms of the effective Hamiltonian the encircled pairs of spins and two single spins (the not encircled ones) are used. The circles on the right hand side of the figure correspond to the range-1 terms in the effective Hamiltonian. These circles are connected by the lines, that correspond to range-2 terms.}
\label{6to4}
\end{figure}

In the last example in this section we consider a ladder of six spins, that one transforms to a plaquette of four spins using a non-uniform blocking (see FIG. \ref{6to4}). Two encircled pairs of spins and two single spins are used to derive the range-1 terms of the effective Hamiltonian. In this case the resulting Hamiltonian will contain also range-4 terms. Our goal here is to show that range-4 terms present in the effective Hamiltonian can be dropped.

\begin{figure}[!ht]
\includegraphics[width=0.45\columnwidth]{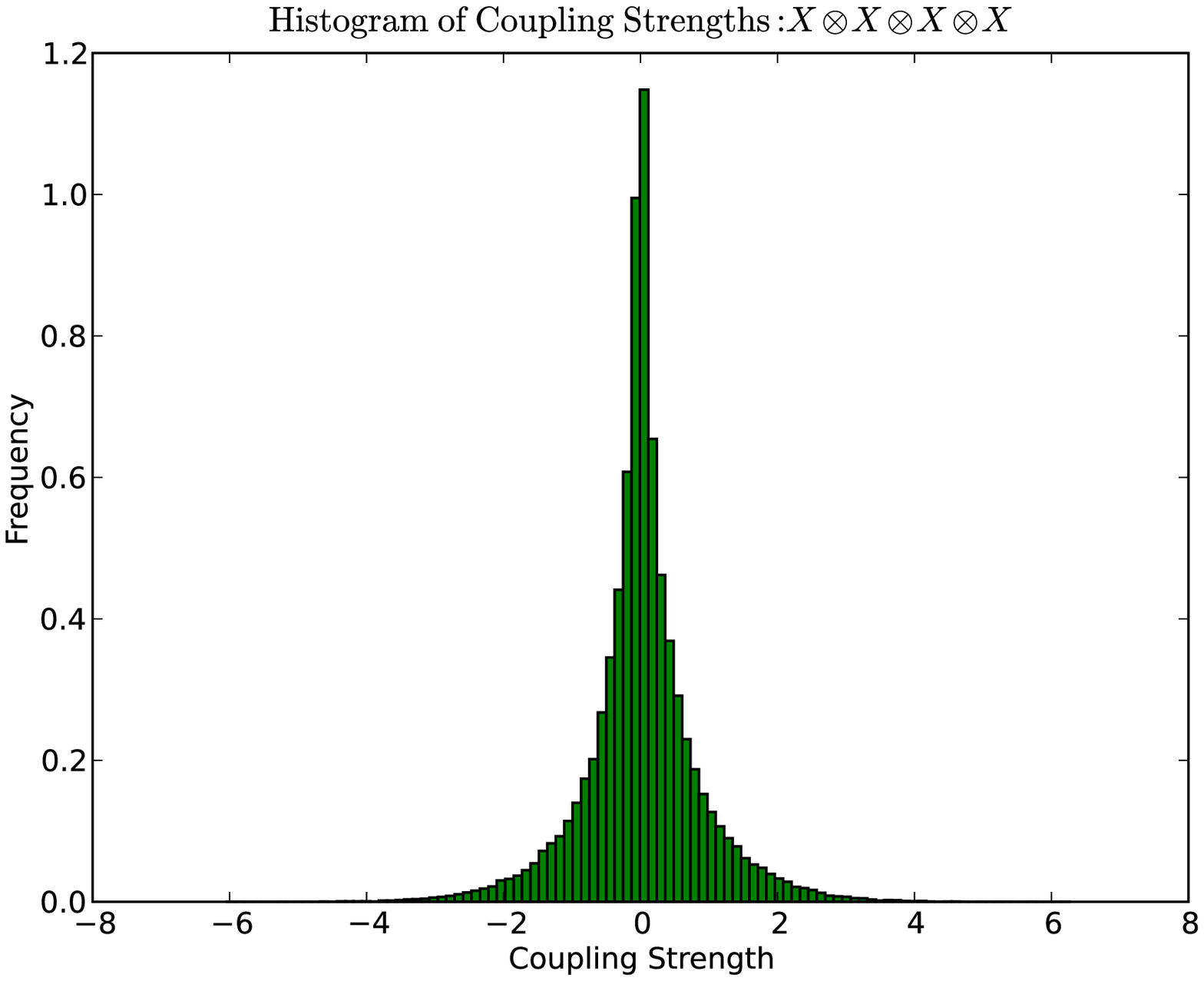}
\includegraphics[width=0.45\columnwidth]{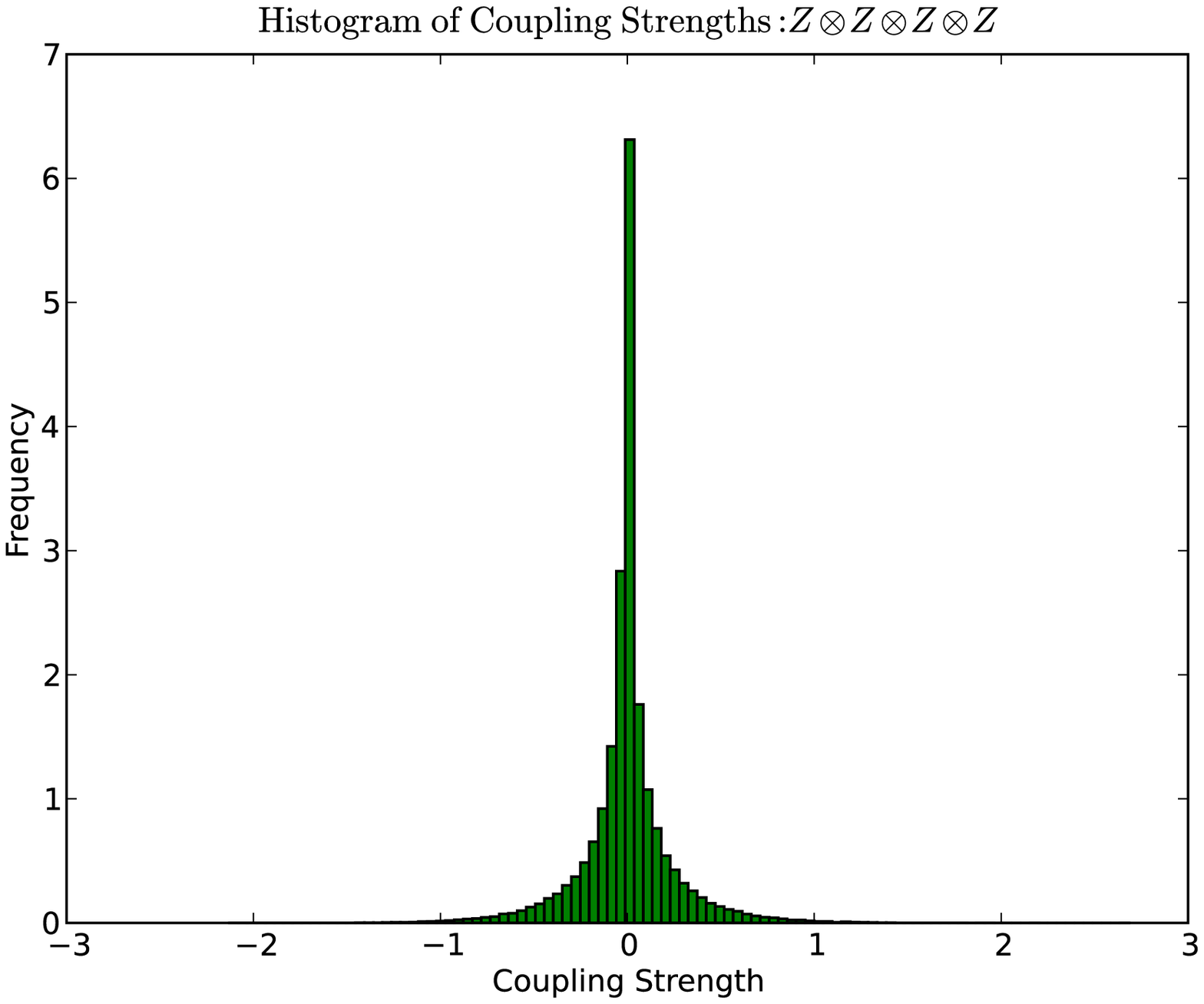}
\caption{(Color online) Renormalization of six spins of the RTFIM using a non-uniform blocking. (Left plot) $XXXX$ plaquette coupling of the range-4 term. $\mean{\sigma_x\otimes\sigma_x\otimes\sigma_x\otimes\sigma_x}=0.024$, $\sigma(X\otimes X\otimes X\otimes X)=0.823$. (Right plot) $ZZZZ$ plaquette coupling of the range-4 term. $\mean{\sigma_z\otimes\sigma_z\otimes\sigma_z\otimes\sigma_z}=0.001$, $\sigma(\sigma_z\otimes\sigma_z\otimes\sigma_z\otimes\sigma_z)=0.240$.}
\label{NNN6to4range4}
\end{figure}

In FIG. \ref{NNN6to4range4} we present statistics for two of the range-4 terms in the effective Hamiltonian, that satisfy the $\mathbb{Z}_2$-symmetry of the Ising model. These terms are $\sigma_x\sigma_x\sigma_x\sigma_x$ and $\sigma_Z\sigma_Z\sigma_Z\sigma_Z$.
The mean value of both distributions can be with a good approximation considered to be zero. The standard deviation is $0.823$ and $0.240$ for the $XXXX$ and $ZZZZ$ term respectively.

\begin{figure}[!ht]
\includegraphics[width=0.45\columnwidth]{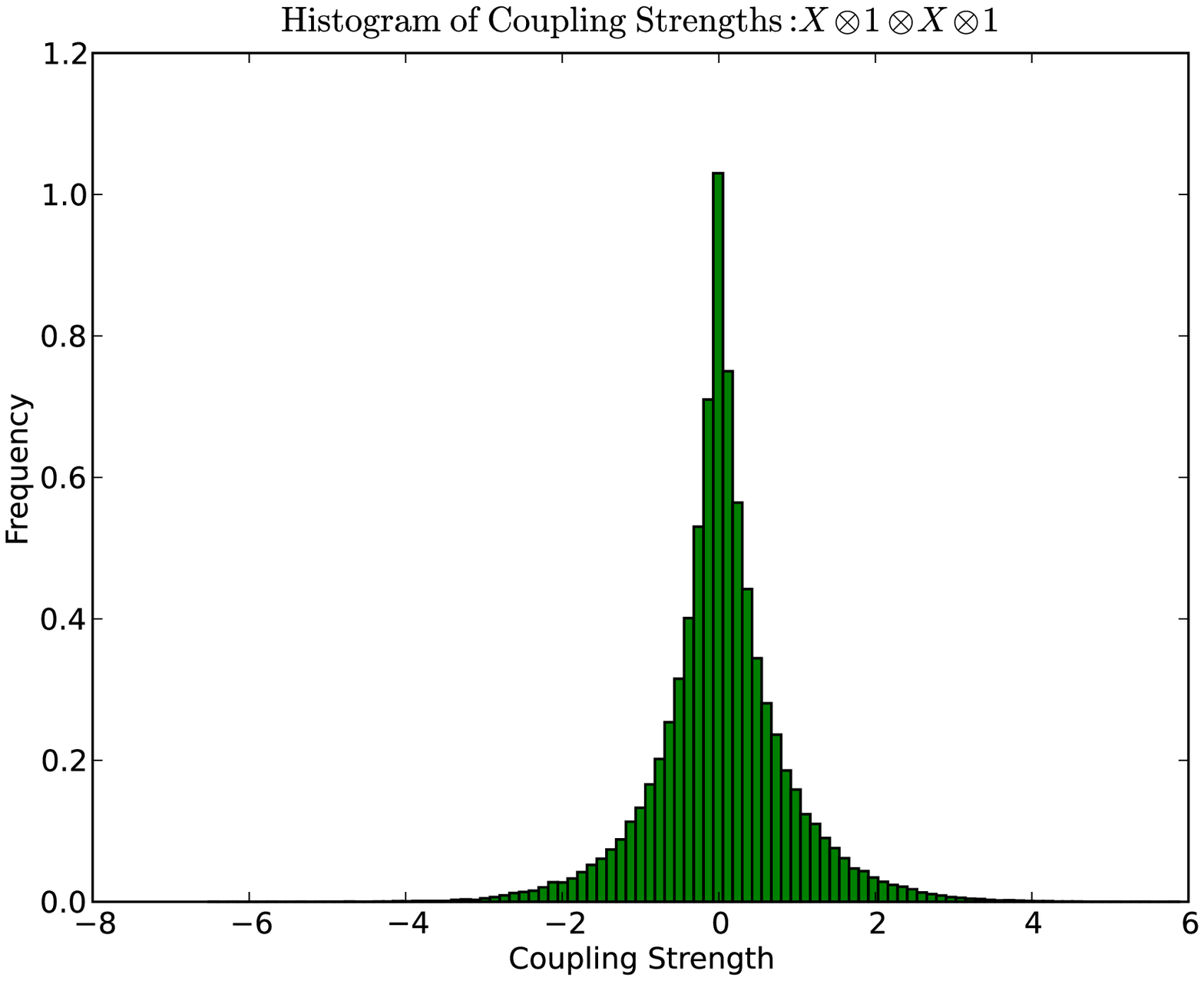}
\includegraphics[width=0.45\columnwidth]{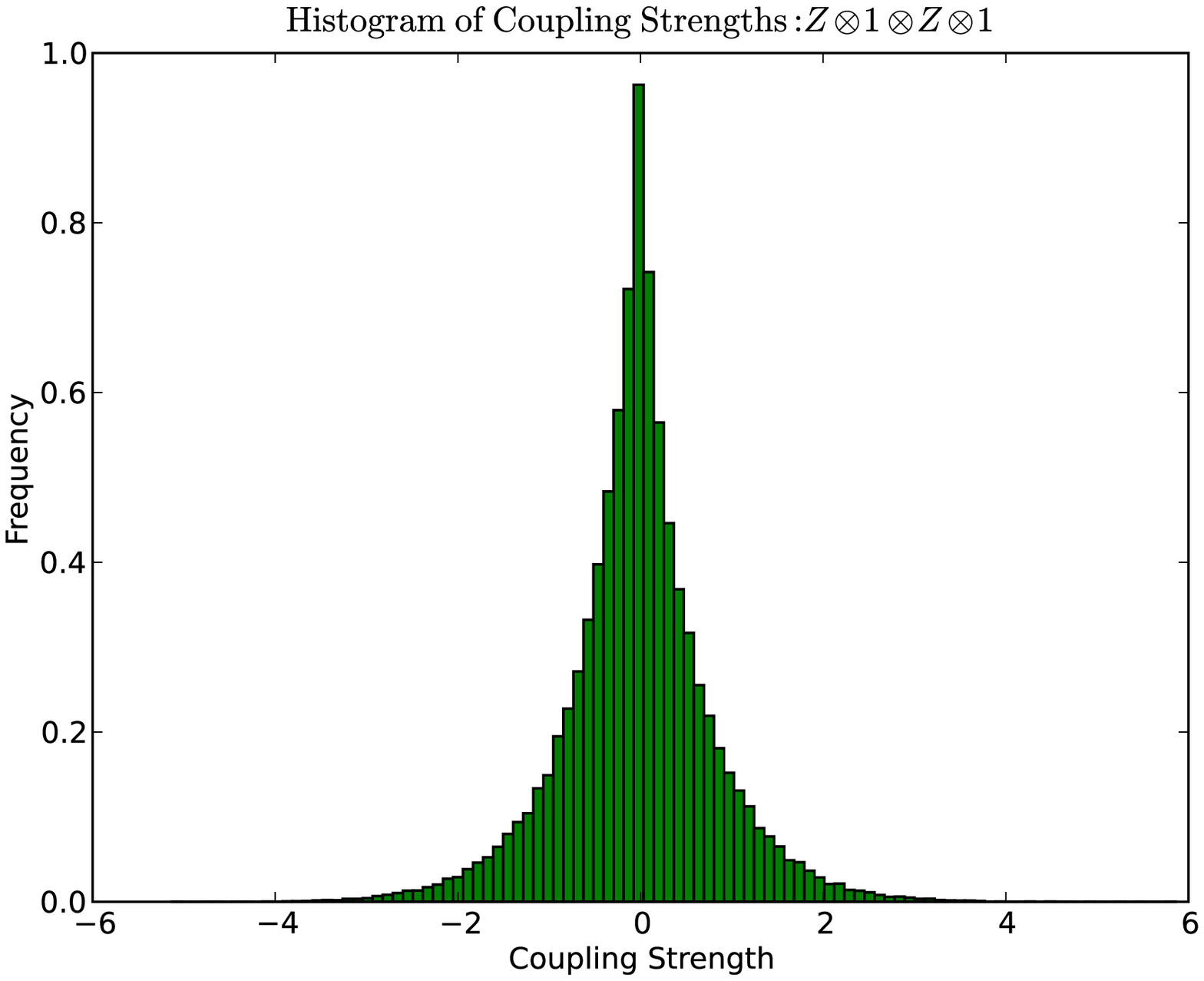}
\caption{(Color online) Renormalization of six spins of the RTFIM using a non-uniform blocking. (Left plot) $XX$ coupling between the first and the third particle. $\mean{\sigma_x\otimes\eins\otimes\sigma_x\otimes\eins}=0.034$, $\sigma(\sigma_x\otimes\eins\otimes\sigma_x\otimes\eins)=0.835$. (Right plot) $ZZ$ coupling between the first and the third particle. $\mean{\sigma_z\otimes\eins\otimes\sigma_z\otimes\eins}=0.032$, $\sigma(\sigma_z\otimes\eins\otimes\sigma_z\otimes\eins)=0.805$.}
\label{NNN6to4range3}
\end{figure}

Finally, we present analogous statistics for the corresponding range-3 terms (FIG. \ref{NNN6to4range3}). These terms correspond to the next nearest neighbor interactions in the renormalized model. We can compare these results with the results of the previous section, where we considered the transformation of the six spin ladder to a three spin chain. The mean value here is $0.034$ for $XX$ and $0.032$ for $ZZ$ interaction. The corresponding standard deviation is $0.835$ and $0.805$ for $XX$ and $ZZ$ interactions respectively.

From the presented examples we conclude that one can apply the non-uniform blocking to perform the renormalization transformation locally in the real space. Our numerical results show that the range-3 and range-4 terms are small and average out and therefore can be neglected in further considerations. Indeed, as it can be seen from our numerics, there are equal number of couplings with negative and positive signs. These contributions cancel each other in average. Since \emph{a priori} no particular distribution of initial couplings was assumed, this fact substantiates the assumption that in the case of appropriate, by means of \cite{Fisher:1995qr}, distribution of couplings the contribution from the long-range interactions to the effective Hamiltonian become negligible.

It could, for example, happen that the encircled pair of spins has a non-degenerate ground state and a double degenerate first excited state. This is true if for example both local magnetic fields are much stronger than the coupling. Such situation is unfavorable for the construction of a range-1 term in the expansion of the effective Hamiltonian, since in every range-1 term we keep two states. This, as we believe, is the main source of errors that cause a rather big variance of the distributions of the strengths of range-3 and range-4 terms presented in this section. Now if we assigned a particular coupling strength to each bond and a particular magnetic field to each spin on the lattice, we would avoid the error, and the all range-3 and range-4 terms would turn exactly to zero, which is illustrated by the fact that all of them have an arbitrarily small mean value. 

In the renormalization transformation, introduced in this paper, one chooses a particular part of a lattice (a ladder), that corresponds to a suitable distribution of couplings and magnetic fields. According to previous numerical evidence, this choice allows to write the resulting Hamiltonian after every renormalization step in the form, which contains only range-1 and range-2 terms
\be
H_{\text{eff}}=  -\sum_{\mean{i,j}} \left( J^z_{ij}\sz{i}\sz{j} + J^x_{ij}\sx{i}\sx{j} + J^y_{ij}\sy{i}\sy{j}  \right) - \sum_i h_i \sx{i}
\ee
with nearest neighbors interactions $J$ and local magnetic field $h^x_i$.
\subsection{Renormalization of the basic constituent of the ladder}

In the last part of this section we investigate the performance of the renormalization transformation applied to a toy model, which is a basic constituent of a ladder. In other words we investigate the basic constituent of the renormalization step, as it is described in section \ref{gen_idea}.
\begin{figure}[!ht]
\includegraphics[width=0.35\columnwidth]{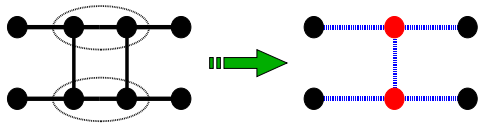}
\caption{(Color online) The basic constituent of the renormalization step: two 4-spin chains, whose central spins are coupled. The renormalization involves two central spins of both chains and results in two new spin-$\halbe$ particles (red circles) and five effective interactions (blue dashed lines).}
\label{tworungs}
\end{figure}

The toy model is presented in the FIG. \ref{tworungs}. Two chains of four spins are coupled such that the central spins form two rungs of a ladder. The renormalization transformation involves two rungs, while the boundary spins keep untouched. The basic renormalized systems consists of six particles that interact as shown in FIG. \ref{tworungs}. Red circles correspond to effective particles that originate from clustering of two central spins of both chains.

\begin{figure}[!ht]
\includegraphics[width=0.9\columnwidth]{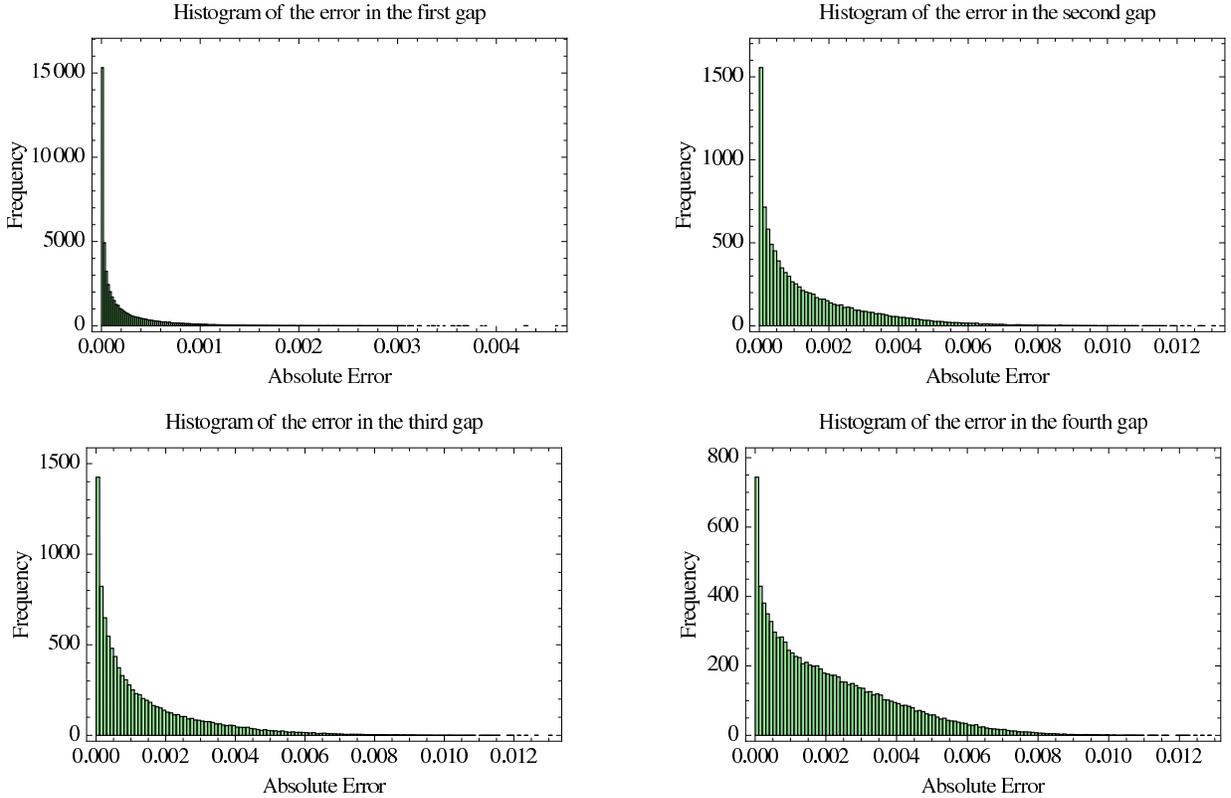}
\caption{(Color online). Renormalization of eight spins of the RTIM using a non-uniform blocking. Histograms of the error in the first eigenvalues between the effective hamiltonian and the exact hamiltonian. The mean values of the errors that appears in the plots are: first gap ($2 ~~10^{-4}$), second gap ($1.4~~ 10^{-3}$), third gap ($1.4~~  10^{-3}$), fourth gap ($2~~ 10^{-3}$)}
\label{tworungsspec}
\end{figure}
In the FIG. \ref{tworungsspec} we compare the spectra of the initial model (with eigenvalues $\lambda^{\text{exact}}_n$) and the model after the renormalization (with eigenvalues $\lambda^{\text{eff}}_n$) and we define the absolute error as $e_n=\frac{|\lambda^{\text{exact}}_n - \lambda^{\text{eff}}_n|}{\lambda^{\text{exact}}_n}$. The absolute error for the first gap is smaller than $10^{-3}$. The error grows slightly, as one considers higher energy levels and is of the order of $6\cdot 10^{-3}$ for the fourth gap. From this observation we conclude that the low energy levels of the initial Hamiltonian are reproduced with a very good accuracy.
\section{Conclusions \& Outlook}
In this paper we have introduced a renormalization transformation for disordered systems on 2D lattices that preserves the geometry of the underlying rectangular lattice. The transformation is done using the real space renormalization group method CORE with non-uniform blocking. We tested the ability of the non-uniform blocking on the random Ising Hamiltonian. Our numerical tests showed that the physics of the ferromagnetic random Ising model can be modeled with nearest neighbor interactions and local magnetic fields in agreement with the conjecture \cite{Motrunich:2000hl}. 

Since Hamiltonian of a spin model can be used to investigate entanglement properties of the model \cite{Guehne:2008gt}, our method provides also a tool for studying the entanglement in 2D disordered quantum spin models. However these investigation will be the subject of forthcoming publication.

Beside that, there are several open problems that can be seen as compendia for future investigations. First, we are going to apply the procedure to other observables. We stress again that the presented technique is a cluster expansion of the Hamiltonian. Therefore in order to be able to calculate the expectation values of other observables (\emph{e.g.} magnetization) we have to know the corresponding cluster expansion. Second, due to the successive nature of the introduced technique (every renormalization step is a sequence of the renormalization transformations, which involve maximally four spins), there is no exponential growth of the time consumption with a systems size. On that account we are going to analyze scaling properties of the introduced method. Third, the way of targeting the region that is to be renormalized depends only on the energy spectrum of the local Hamiltonian of two spins. Thus the method is applicable to any type of Hamiltonian with nearest neigbbors interactions, whose two lowest lying energy states are separated by a gap from the rest of the spectrum, depending on its intrinsic parameters. It is also natural to consider spin models with a higher dimensionality.

Furthermore we argue that there is a rigorous analytical form of the introduced renormalization transformation and that the renormalization flow has a certain fixed point.

Another point of future work is to generalize the method to lattices of non-rectangular geometries. One can consider a variety of different lattice structures. The renormalization procedure will depend on the particular form of the lattice and must be considered anew in every particular case.


We close by mentioning that our method offers itself to go beyond the usual randomness and investigate models possessing a spin glass phase.
\section{Acknowledgements}

The authors acknowledge comments and suggestions that appeared in several conversations with A. Auerbach, L. F. Cugliandolo, M. A. Martin-Delgado, F. Verstraete.

This work was supported by the Austrian Science Foundation (FWF), and the European Union (OLAQUI,SCALA,QICS).

\bibliographystyle{unsrt}

\end{document}